\documentclass{iaus}
\usepackage{graphicx}
\usepackage{mathptmx}
\usepackage{natbib}

\newcommand{\teff}{$T_{\!\mbox{\scriptsize\em eff}}$}

\newcommand{\teffq}{$T_{\!\mbox{\scriptsize \em eff}}^4$}
\newcommand{\zsun}{$Z_\odot$}

\title[Extragalactic Stellar Astronomy] 
{Extragalactic Stellar Astronomy with the \\ Brightest Stars in the Universe}

\author[Rolf Kudritzki, Miguel A.~Urbaneja, Fabio Bresolin \& Norbert Przybilla]   
{Rolf Kudritzki$^1$, Miguel A.~Urbaneja$^2$, Fabio Bresolin$^3$ 
 \and Norbert Przybilla$^4$}

\affiliation{$^1$Institute for Astronomy, University of Hawaii \\ 2680 Woodlawn
Drive, Honolulu, HI 96822, USA \\ email: {\tt kud@ifa.hawaii.edu} \\[\affilskip]
$^2$ Institute for Astronomy, University of Hawaii \\ email: {\tt urbaneja@ifa.hawaii.edu} \\[\affilskip]
$^3$ Institute for Astronomy, University of Hawaii \\ email: {\tt bresolin@ifa.hawaii.edu} \\[\affilskip]
$^4$Dr. Remeis-Sternwarte Bamberg, Erlangen University\\ Sternwartstr. 7, D-96049 Bamberg, Germany\\ email: {\tt przybilla@sternwarte.uni-erlangen.de}}

\pubyear{2008}
\volume{250}  
\pagerange{1--12}
\jname{Massive Stars as Cosmic Engines}
\editors{F. Bresolin, P.A. Crowther \& J. Puls, eds.}
\begin{document}

\maketitle

\begin{abstract}
A supergiants are objects in transition from the blue to the red (and
vice versa) in the uppermost HRD. They are the intrinsically brightest
"normal" stars at visual light with absolute visual magnitudes up to -9.
They are ideal to study young stellar populations in galaxies beyond the
Local Group to determine chemical composition and evolution, interstellar
extinction, reddening laws and distances. We discuss most recent
results on the quantitative spectral analysis of such objects in galaxies
beyond the Local Group based on medium and low resolution spectra obtained 
with the ESO VLT and Keck. We describe the analysis method including
the determination of metallicity and metallicity gradients. A new method 
to measure accurate extragalactic distances based on the stellar gravities 
and effective temperatures is presented, the flux weighted 
gravity - luminosity relationship (FGLR). The FGLR is a purely spectroscopic 
method, which overcomes the untertainties introduced by interstellar 
extinction and variations of metallicity, which plague all photometric 
stellar distance determination methods. We discuss the perspectives of 
future work using the giant ground-based telescopes of the next 
generation such as the TMT, the GMT and the E-ELT.
\end{abstract}

\firstsection 
\section{Introduction}

It has long been the dream of stellar astronomers to study individual stellar 
objects in distant galaxies to obtain detailed spectroscopic information about 
the star formation history and chemodynamical evolution of galaxies and to
determine accurate distances based on the determination of stellar 
parameters and interstellar reddening and extinction. At the first glance, 
one might think that the most massive and, therefore, most luminous stars with 
masses higher than 50 $M_{\odot}$ are ideal for this purpose. However, because 
of their very strong stellar winds and mass-loss these objects keep very hot 
atmospheric temperatures throughout their life and, thus, waste most of their 
precious photons in the extreme ultraviolet. As we all know, most of these UV 
photons are killed by dust absorption in the star forming regions, where these 
stars are born, and the few which make it to the earth can only be observed with 
tiny UV telescopes in space such as the HST or FUSE and are not accessible to 
the giant telescopes on the ground. 

Thus, one learns quickly that the most promising objects for such studies are massive 
stars in a mass range between 15 to 40 $M_{\odot}$ in the short-lived evolutionary
phase, when they leave the hydrogen main-sequence and cross the HRD in a few 
thousand years as blue supergiants of late B and early A spectral type. Because of 
the strongly reduced absolute value of bolometric correction when evolving towards
smaller temperature these objects increase their brightness in visual light and 
become the optically brightest ``normal'' stars in the universe with absolute visual 
magnitudes up to $M_{V} \cong -9.5$ rivaling with the integrated light brightness 
of globular clusters and dwarf spheroidal galaxies. These are the ideal stellar 
objects to obtain accurate quantitative information about galaxies.

\section{Studies in the Milky Way and Local Group}

There has been a long history of quantitative spectroscopic studies of these 
extreme objects. In a pioneering and comprehensive paper on Deneb 
\citet{groth61} was the first to obtain stellar parameters and detailed 
chemical composition. This work was continued by Wolf (1971, 1972, 1973) 
in studies of A supergiants in the Milky Way and the Magellanic 
Clouds. \cite{kud73} using newly developed NLTE model atmospheres found 
that at the low gravities and the correspondingly low electron densities of these
objects effects of departures from LTE can become extremely important. With strongly
improved model atmospheres \cite{venn95a}, \cite{venn95b} and \cite{aufdenb02}
continued these studies in the Milky Way. Most recently, \cite{przybilla06} and 
\cite{schiller08} used
very detailed NLTE line formation calculations including ten thousands of lines in NLTE
(see also \cite{przybilla00}, \cite{przybilla01a}, \cite{przybilla01b},
\cite{przybilla01}, \cite{przybilla02}) to determine stellar parameters and
abundances with hitherto unkown precision (\teff\/ to $\le 2$\%, $log~g$ to $\sim 0.05$ dex,
individual metal abundances to $\sim 0.05$ dex).
At the same time, utelizing the power of the new 8m to 10m class telescopes, high resolution 
studies of A supergiants in many Local Group galaxies  were carried out by \cite{venn99} (SMC),
\cite{mccarthy95} (M33), \cite{mccarthy97} (M31), \cite{venn00} (M31),
\cite{venn01} (NGC 6822), \cite{venn03} (WLM), and \cite{kaufer04} (Sextans A) 
yielding invaluable information about the stellar chemical composition in these galaxies.
In the research field of massive stars, these studies have so far provided the most 
accurate and most comprehensive information about chemical composition and have been used 
to constrain stellar evolution and the chemical evolution of their host galaxies.

\section{The Challenging Step beyond the Local Group}

The concept to go beyond the Local Group and to study A supergiants by means of 
quantitative spectroscopy in galaxies out to the Virgo cluster has been first
presented by \cite{kud95} and \cite{kud98}. Following-up on this idea,
\cite{bresolin01} and \cite{bresolin02} used the VLT and FORS at 5 $\AA$ resolution
for a first investigation of blue supergiants in NGC 3621 (6.7 Mpc) and NGC 300 (1.9 Mpc). They were able
to demonstrate that for these distances and at this resolution spectra of sufficient S/N 
can be obtained allowing for the quantitative determination of stellar parameters 
and metallicities. \cite{kud03} extended this work and showed that stellar 
gravities and temperatures determined from the spectral analysis can be used to 
determine distances to galaxies by using the correlation between absolute bolometric
magnitude and flux weighted gravity $g_{F} = g/$\teffq\/ (FGLR).
However, while these were encouraging steps towards the use of A supergiants as quantitative
diagnostic tools of galaxies beyond the Local Group, the work presented in these papers
had still a fundamental deficiency. At the low resolution of 5 $\AA$ it is not possible
to use ionization equilibria for the determination of \teff\/ 
in the same way as in the high resolution 
work mentioned in the previous paragraph. Instead, spectral types were determined and
a simple spectral type - temperature relationship as obtained for the Milky Way was 
used to determine effective temperatures and then gravities and metallicities. Since
the spectral type - \teff\/ relationship must depend on metallicity (and also gravities),
the method becomes inconsistent as soon as the metallicity is significantly different from
solar (or the gravities are larger than for luminosity class Ia) and may lead to inaccurate
stellar parameters. As shown by  \cite{evans03}, the uncertainties introduced in this 
way could be significant and would make it impossible to use the FGLR for distance 
determinations. In addition, the metallicities derived might be unreliable. This posed
a serious problem for the the low resolution study of A supergiants in distant galaxies.

This problem was overcome only very recently by \cite{kud08} (herafter KUBGP), who provided the first 
self-consistent determination of stellar parameters and metallicities for A supergiants 
in galaxies beyond the Local Group based on the detailed quantitative model atmosphere 
analysis of low resolution spectra. They applied their new method on 24 supergiants of 
spectral type
B8 to A5 in the Scultor Group spiral galaxy NGC 300 (at 1.9 Mpc distance) and obtained 
temperatures, gravities, metallicities, radii, luminosities and masses. The spectroscopic 
observations were obtained with FORS1 at the ESO VLT in multiobject 
spectroscopy mode. In addition, ESO/MPI 2.2m WFI and HST/ACS photometry was used. The observations 
were carried out within the framework of the Araucaria Project (\cite{gieren05b}).
In the following we discuss the analysis method and the results of this pilot study. 

\begin{figure}[b]
\begin{minipage}{7cm}
\resizebox{\hsize}{!}
 {\includegraphics[angle=90,width=6cm]{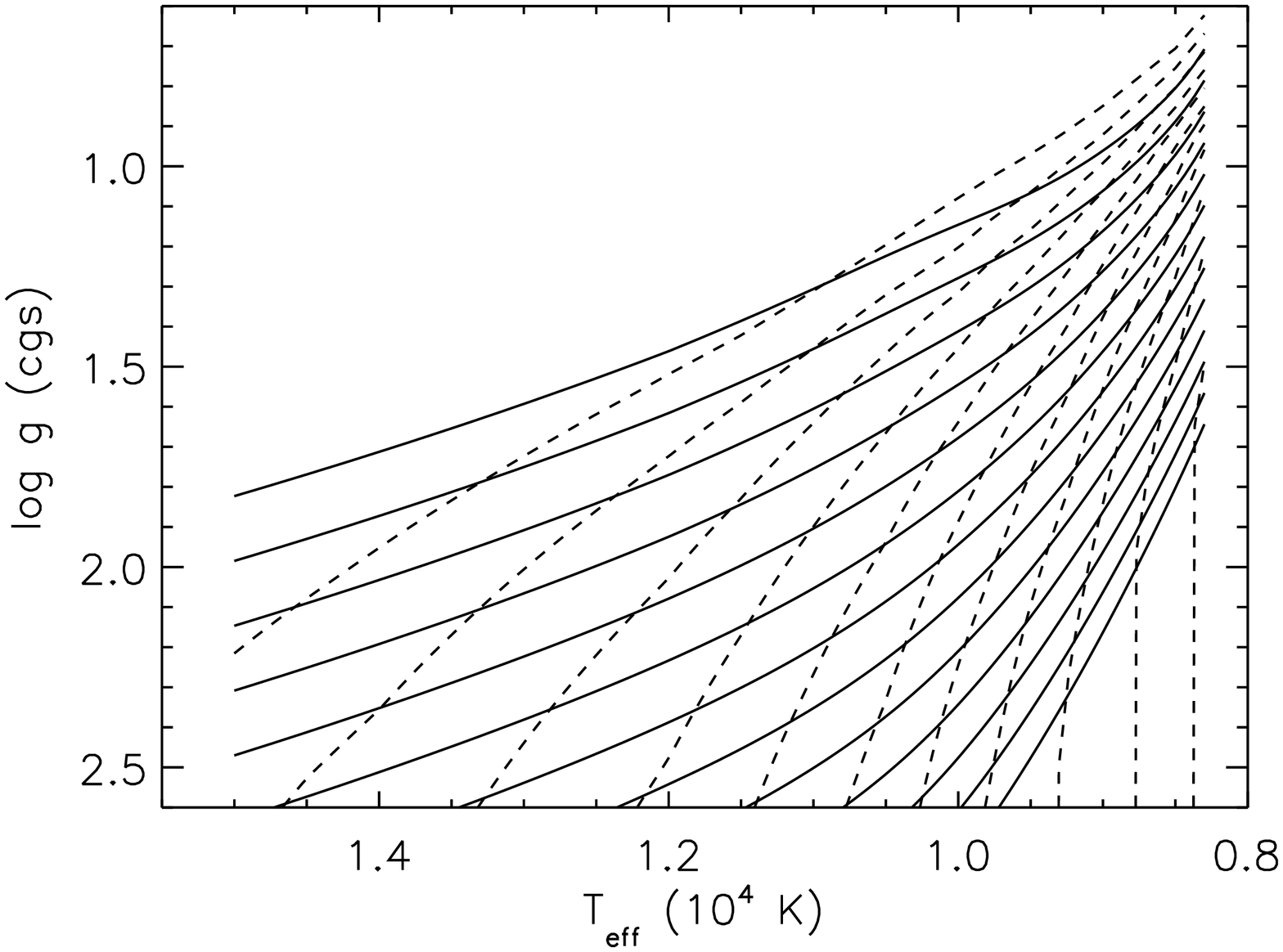}} 
\end{minipage}
\hfill
\begin{minipage}{7cm}
\resizebox{\hsize}{!}
 {\includegraphics[angle=90,width=6cm]{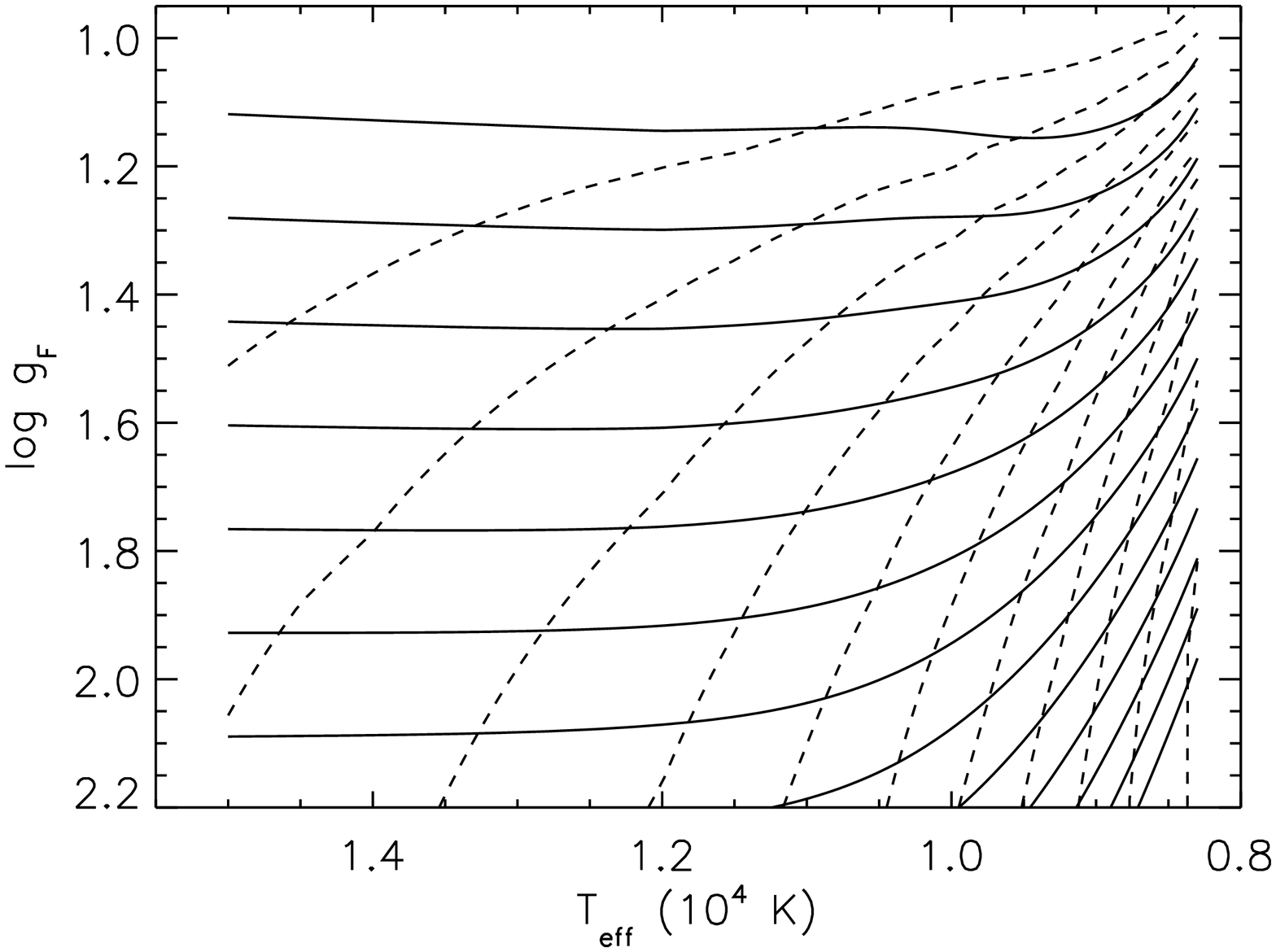}} 
\end{minipage}
 \caption{{\bf Left:}Isocontours of H${\delta}$ equivalent widths (solid) and Balmer 
jump $D_{B}$ (dashed) in the $(log~g, log$~\teff) plane. H${\delta}$ isocontours 
start with 1 \AA~equivalent width and increase in steps of 0.5 \AA. $D_{B}$ 
isocontours start with 0.1 dex and increase by 0.1 dex. 
{\bf Right:} Same as left but for the flux weighted gravity $log~g_{F}$ 
instead of gravity $log~g$.
Note that this diagram is independent of metallicity, since both the strengths 
of Balmer lines and the Balmer jump depend only very weakly on metallicity.}
   \label{fig1}
\end{figure}

\section{A Pilot Study in NGC300 - Analysis Method}

For the quantitative analysis of the spectra KUBGP use the same combination
of line blanketed model atmospheres and very detailed NLTE line formation calculations
as \citet{przybilla06} in their high signal-to-noise and high spectral resolution
study of galactic A-supergiants, which reproduce the observed normalized spectra and 
the spectral energy distribution, including the Balmer jump, extremely well. 
They calculate an extensive, comprehensive and dense grid of model atmospheres and NLTE 
line formation covering the potential full parameter range of all the objects in 
gravity ($log~g$ = 0.8 to 2.5), effective temperature (\teff = 8300 to 15000K) and 
metallicity ([Z] = log{Z/\zsun} = -1.30 to 0.3). The total grid comprises more than 6000 
models.

\begin{figure}[b]
\begin{minipage}{7cm}
\resizebox{\hsize}{!}
   {\includegraphics[width=6cm]{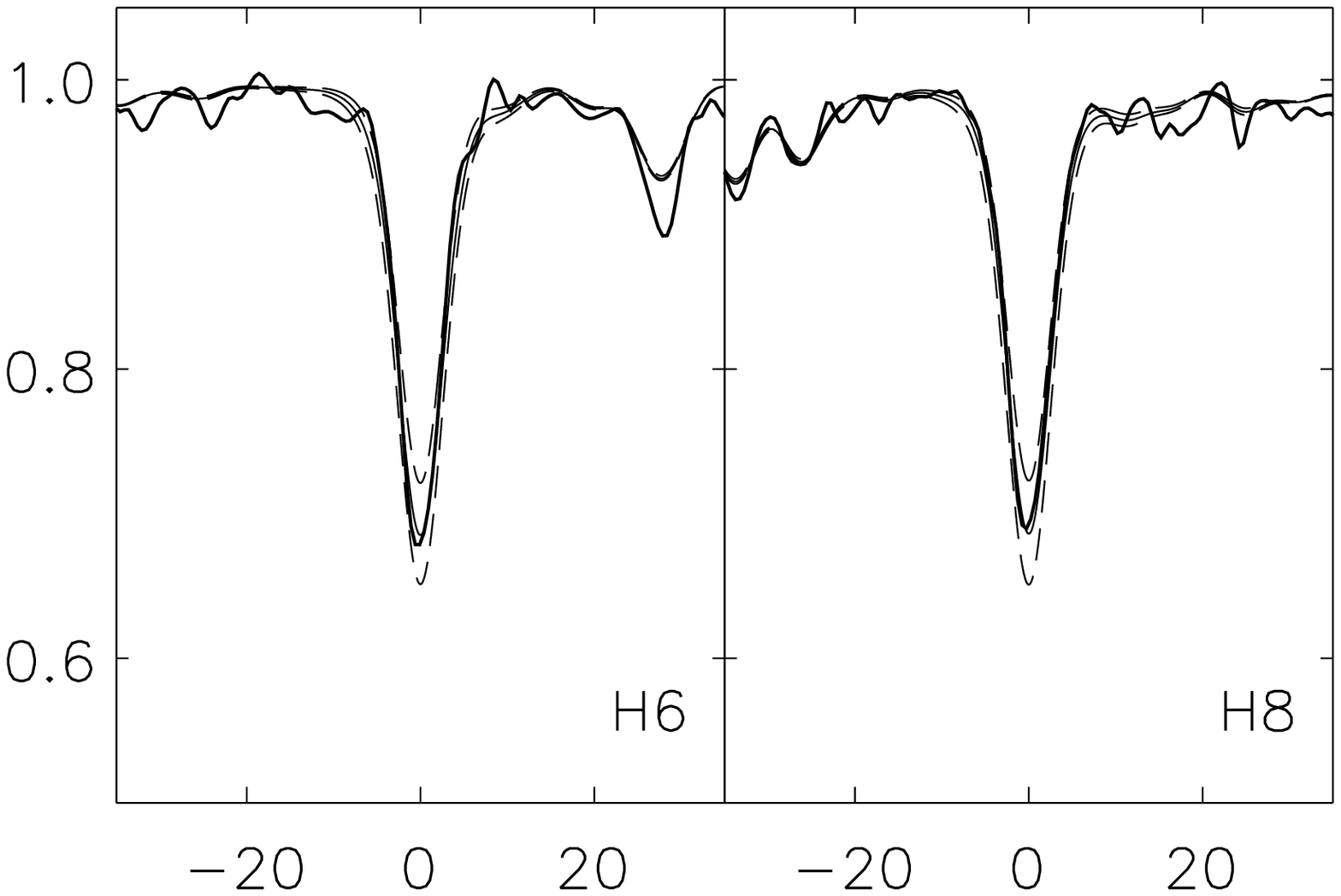}}
\end{minipage}
\hfill
\begin{minipage}{7cm}
   \resizebox{\hsize}{!}
   {\includegraphics[angle=90,width=6cm]{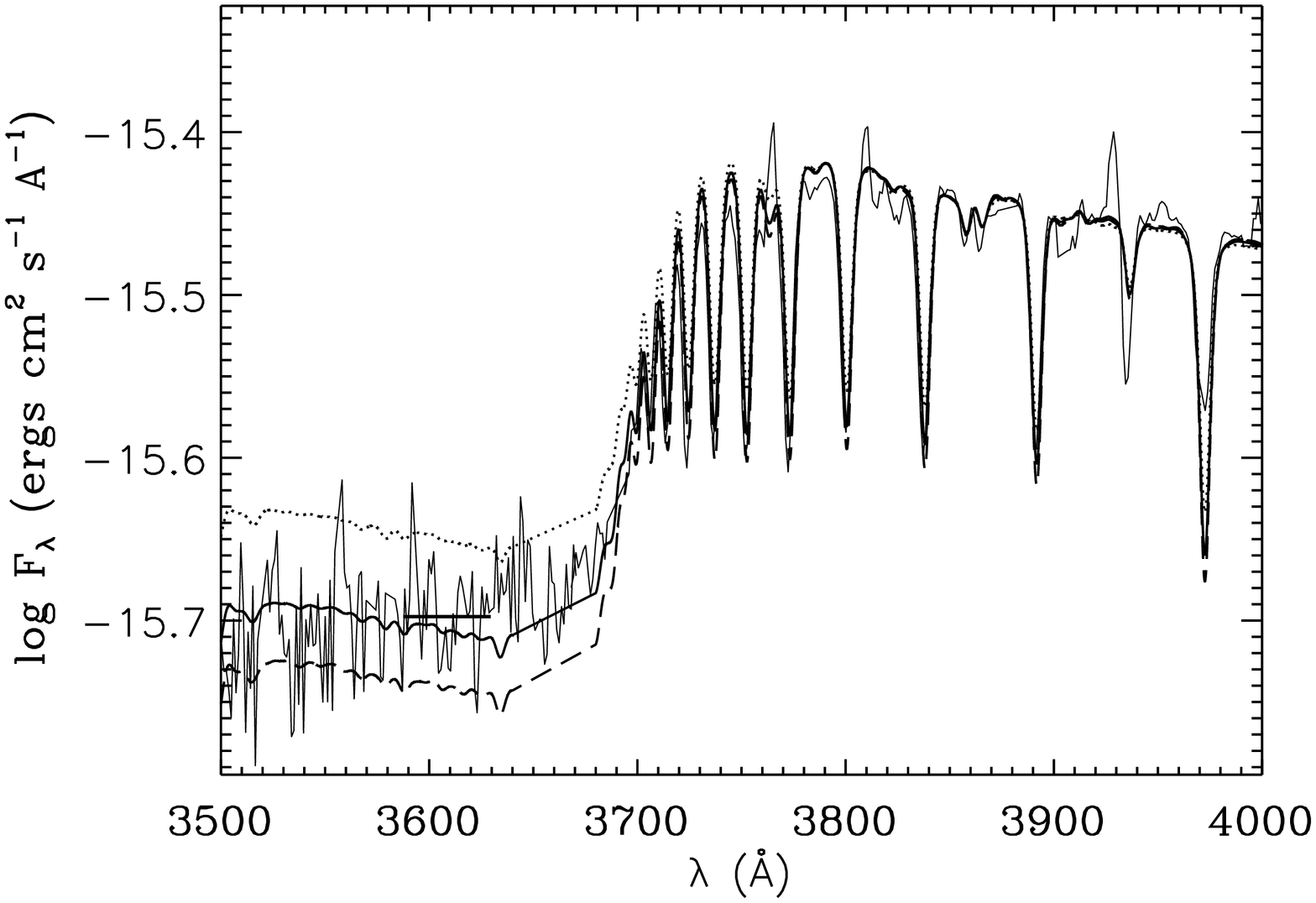}}
\end{minipage} 
\caption{{\bf Left:}  Model atmosphere fit of two observed Balmer lines of 
NGC300 target No. 21 of KUBGP for \teff\/ = 10000~K and $log~g$ = 1.55 (solid). 
Two additional models with same \teff\/ but $log~g$ = 1.45 and 1.65, 
respectively, are also shown (dashed). {\bf Right:}  Model atmosphere fit of 
the observed Balmer jump of the same target for 
\teff\/ = 10000~K and $log~g$ = 1.55 (solid). Two additional models with the same $log~g$
but \teff\/ = 9750~K (dashed) and 10500~K (dotted) are also shown. The horizontal bar at 
3600~\AA~represents the average of the flux logarithm over this wavelength interval, which 
is used to measure D${_B}$.}
   \label{fig2}
\end{figure}

The analysis of the each of the 24 targets in NGC 300 proceeds in three steps. First, 
the stellar 
parameters (\teff\/ and $log~g$) are determined together with interstellar reddening 
and extinction, then the metallicity is determined  and finally, assuming a distance to 
NGC 300, stellar radii, luminosities and masses are obtained. For the first step,
a well established method to obtain the stellar parameters of supergiants of late B to 
early A spectral type is to use ionization equilibria of weak metal lines (OI/II; MgI/II;
NI/II etc.) for the determination of effective temperature \teff\/ and the Balmer lines 
for the gravities $log~g$. However, at the low resolution
of 5 \AA~the weak spectral lines of the neutral species disappear in the noise of the 
spectra and an alternative technique is required to obtain temperature information. 
KUBGP confirm the result by \cite{evans03} that a simple application of a spectral 
type - effective temperature relationship does not work because of the degeneracy of 
such a relationship with metallicity. Fortunately, a way out of this dilemma is the 
use of the spectral energy distributions (SEDs) and here, in particular of the Balmer 
jump $D_B$. While the observed photometry from B-band to I-band is 
used to constrain the interstellar reddening, $D_B$ turns out to be a reliable 
temperature diagnostic, as is indicated by Fig.\,\ref{fig1}. A simultaneous fit of the 
Balmer lines and the Balmer jump allows to constrain effective temperature and gravity 
independent of assumptions on metallicity. Fig.\,\ref{fig2} demonstrates the sensitivity 
of the Balmer lines and the Balmer jump to gravity and effective temperature, respectively.
At a fixed temperature the Balmer lines allow for a determination of $log~g$ within 0.05 
dex uncertainty, whereas the Balmer jump at a fixed gravity yields a temperature 
uncertainty of 2 percent. However, since the isocontours in  Fig.\,\ref{fig1} are not 
orthogonal, the maximum errors of \teff ~and $log~g$ are 5 percent and 0.2 dex, respectively.
These errors are significantly larger than for the analysis of high resolution spectra, but 
they still allow for an accurate determination of metallicity and distances.

The accurate determination of \teff\/ and $log~g$ is crucial for the use of
A supergiants as distance indicators using the relationship between absolute 
bolometric magnitude $M_{bol}$ and flux weighted gravity $log~g_{F}$ defined as

\begin{equation}
        log~g_{F} = log~g - 4log T_{\!\mbox{\scriptsize\em eff,4}}
   \end{equation}

where $T_{\!\mbox{\scriptsize\em eff,4}} =  T_{\!\mbox{\scriptsize\em eff}}$/10000K
(see \citealt{kud03}). 
The relatively large uncertainties obtained with this fit 
method may casts doubts whether $log~g_{F}$ can be obtained accurately enough.
Fortunately, the non-orthogonal behaviour of the fit curves in the left part 
of Fig.\,\ref{fig1} 
leads to errors in \teff\/ and $log~g$, which are correlated in a way 
that reduces the uncertainties of $log~g_{F}$. This is demonstrated in right 
part of Fig.\,\ref{fig1}, which shows the corresponding fit curves of the Balmer lines
and D$_{B}$ in the $(log~g_{F}, log$~\teff) plane. As a consequence, much smaller 
uncertainties are obtained for $log~g_{F}$, namely $\pm$ 0.10 dex.
KUBGP give a detailed discussion of physical reason behind this.

\begin{figure}[b]
\begin{center}
 \includegraphics[width=8cm]{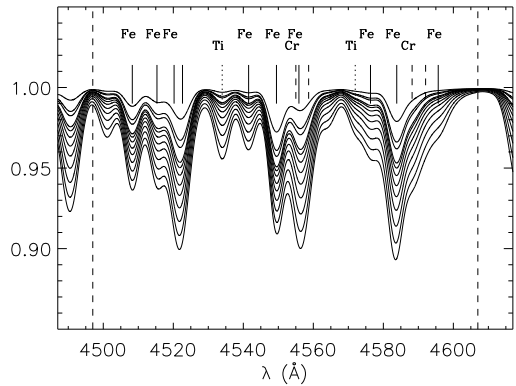} 
 \caption{Synthetic metal line spectra calculated for the stellar parameters of target 
No.21 as a function of metallicity in the spectral window from 4497~\AA~to 4607~\AA. 
Metallicities range from [Z] = -1.30 to 0.30, as described in the text. The dashed 
vertical lines give the edges of the spectral window as used for a determination of 
metallicity.}
   \label{fig3}
\end{center}
\end{figure}

\begin{figure}[b]
\begin{minipage}{7cm}
\resizebox{\hsize}{!}
   {\includegraphics[width=6cm]{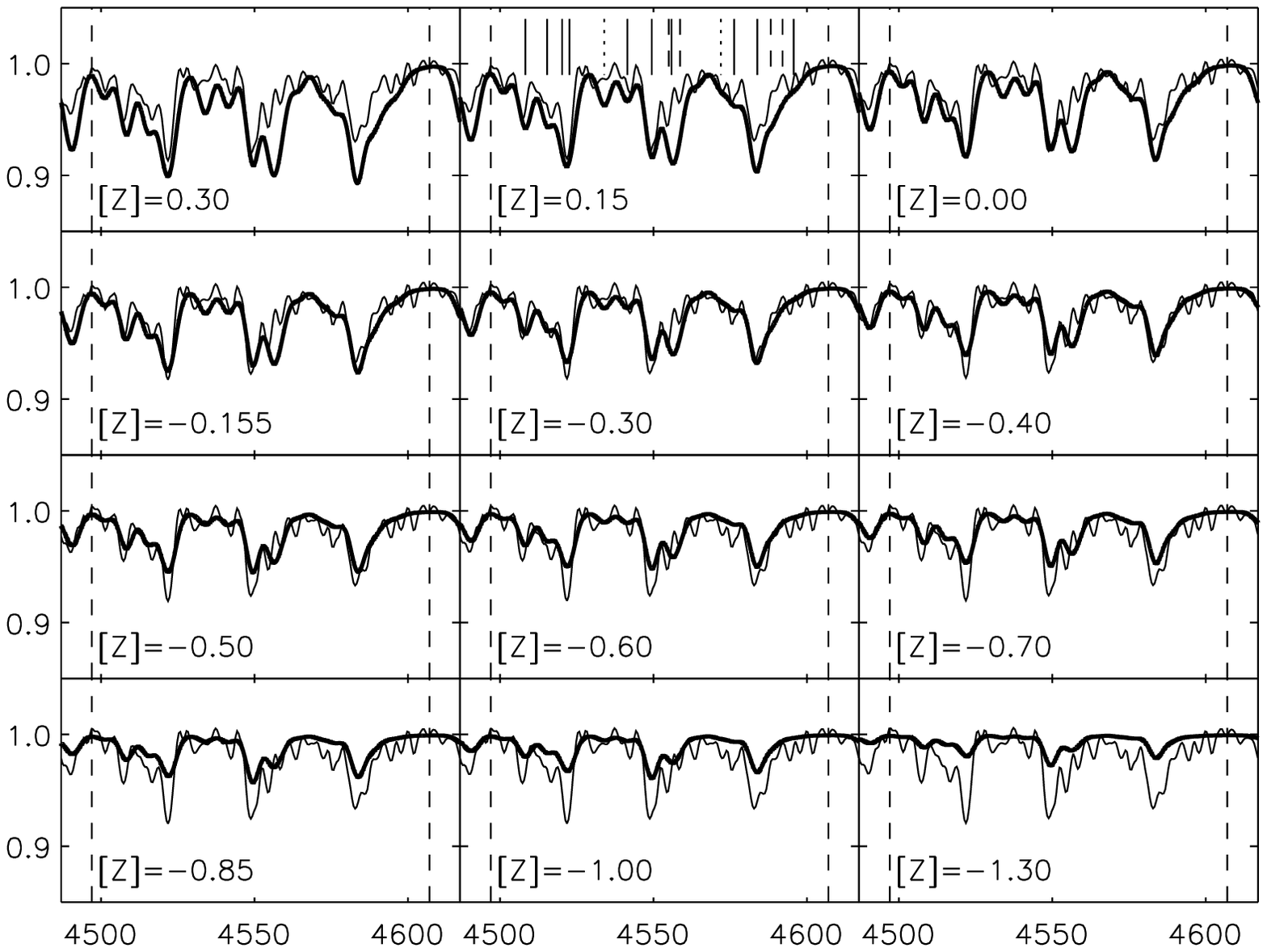}}
\end{minipage}
\hfill
\begin{minipage}{7cm}
   \resizebox{\hsize}{!}
   {\includegraphics[width=6cm]{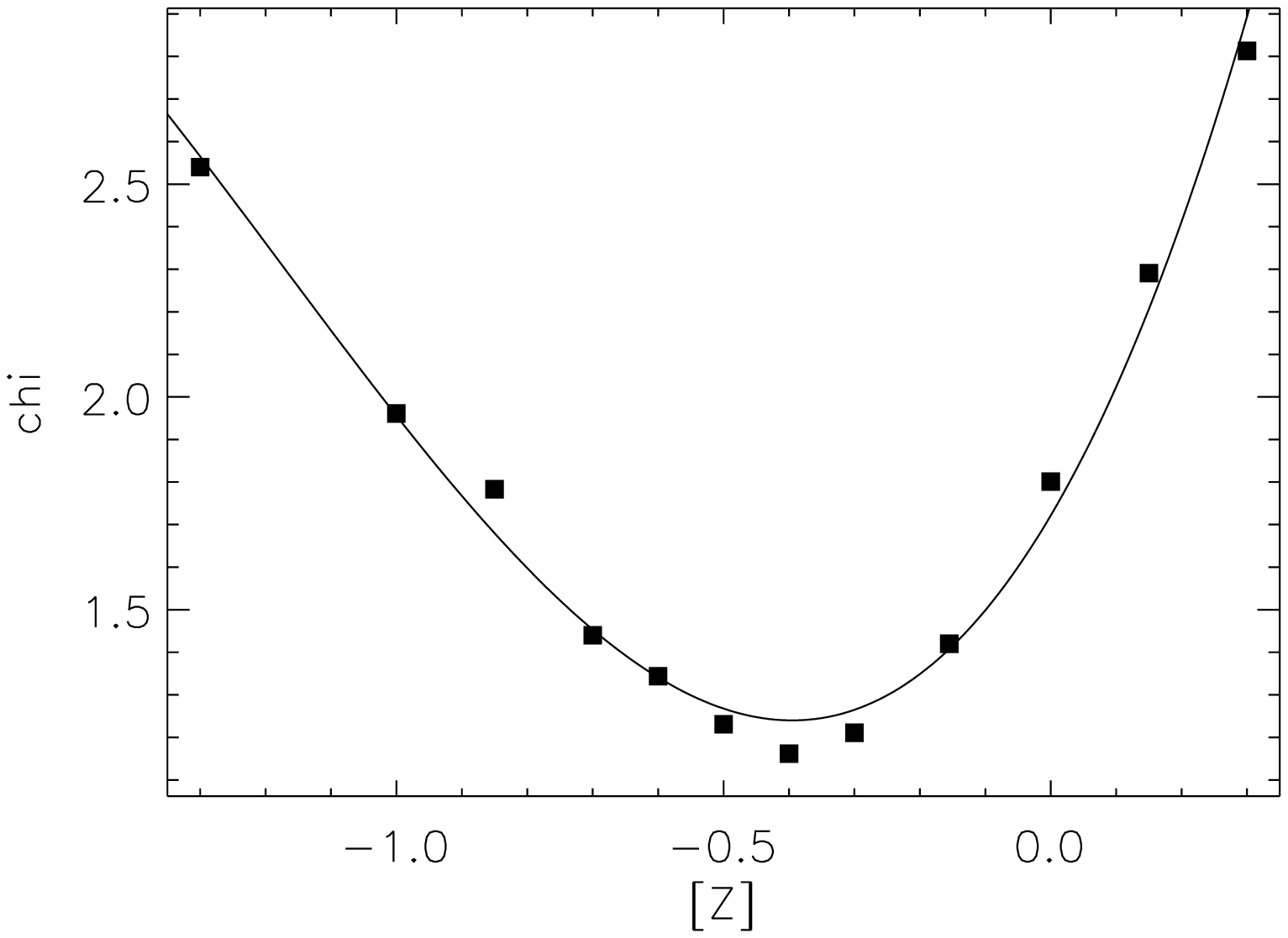}}
\end{minipage} 
\caption{{\bf Left:} Observed spectrum of target No. 21 for the same spectral window as 
Fig.\,\ref{fig3} overplotted by the same synthetic spectra for each metallicity 
separately. {\bf Right:} $\chi ([Z])$ as obtained from the comparison of observed and 
calculated spectra. The solid curve is a third order polynomial fit.}
   \label{fig4}
\end{figure}

Knowing the stellar atmospheric parameters \teff\/ and $log~g$ KUBGP are able to 
determine stellar metallicities by fitting the metal lines with their 
comprehensive grid of line formation calculations. The fit procedure proceeds 
in several steps. First, spectral windows are defined, for which a good definition 
of the continuum is possible and which are relatively undisturbed by flaws in 
the spectrum (for instance caused by cosmic events) or interstellar emission 
and absorption. A typical spectral window used for all targets is the 
wavelength interval 4497~\AA\/  $\le \lambda \le$ 4607~\AA. Fig.\,\ref{fig3} shows 
the synthetic spectrum calculated for the atmopsheric parameters of target 
No. 21 (the previous example) and for all the metallicities 
of the grid ranging from -1.30 $\le$ [Z] $\le$ 0.30. It is very obvious that the 
strengths of the metal line features are a strong function of metallicity. In 
Fig.\,\ref{fig4} the observed spectrum of target No. 21 in this spectral 
region is shown overplotted by the synthetic spectrum for each metallicity. Separate 
plots are used for each metallicity, because the optimal relative normalization of the observed and 
calculated spectra is obviously metallicity dependent. This problem is addressed by 
renormalizing the observed spectrum for each metallicity so that the synthetic 
spectrum always intersects the observations at the same value at the two edges of 
the spectral window (indicated by the dashed vertical lines).  
The next step is a pixel-by-pixel comparison of calculated and normalized observed 
fluxes for each metallicity and a calculation of a $\chi^{2}$-value. The minimum 
$\chi ([Z])$ as a function of [Z] is then used to determine the metallicity. This 
is also shown in Fig.\,\ref{fig4}. Application of the same method on different spectral
windows provides additional independent information on 
metallicity and allows to determine the average metallicity 
obtained from all windows. A value of is [Z] = -0.39 with a very small dispersion of only 
0.02 dex. However, one also need to consider the 
effects of the stellar parameter uncertainties on the metallicity determination. This 
is done by applying the same correlation method for [Z] for models at the extremes 
of the error box for \teff\/ and $log~g$. This increases the uncertainty of [Z] 
to $\pm$ 0.15 dex, still a very reasonable accuracy of the abundance determination.

The fit of the observed photometric fluxes with the model atmosphere fluxes was 
used to determine interstellar reddening E(B-V) and extinction A$_{V}$ = 3.1 E(B-V). 
Simultaneously, the fit also yields the stellar angular diameter, which provides 
the stellar radius, if a distance is adopted. Alternatively, for the stellar 
parameters (\teff\/, $log~g$, [Z]) determined through the spectral analysis
the model atmospheres also yield bolometric corrections BC, which we use to 
determine bolometric magnitudes. These bolometric magnitudes then also allow us to 
compute radii. The radii determined with these two different methods agree within
a few percent. \cite{gieren05} in their multi-wavelength study of a large sample 
of Cepheids in NGC 300 including the near-IR have determined a new distance 
modulus m-M = 26.37 mag, which corresponds to a distance of 1.88 Mpc. KUBGP have 
adopted these values to obtain the radii and absolute magnitudes. 

\begin{figure}[b]
\begin{minipage}{7cm}
\resizebox{\hsize}{!}
   {\includegraphics[width=6cm,angle=90]{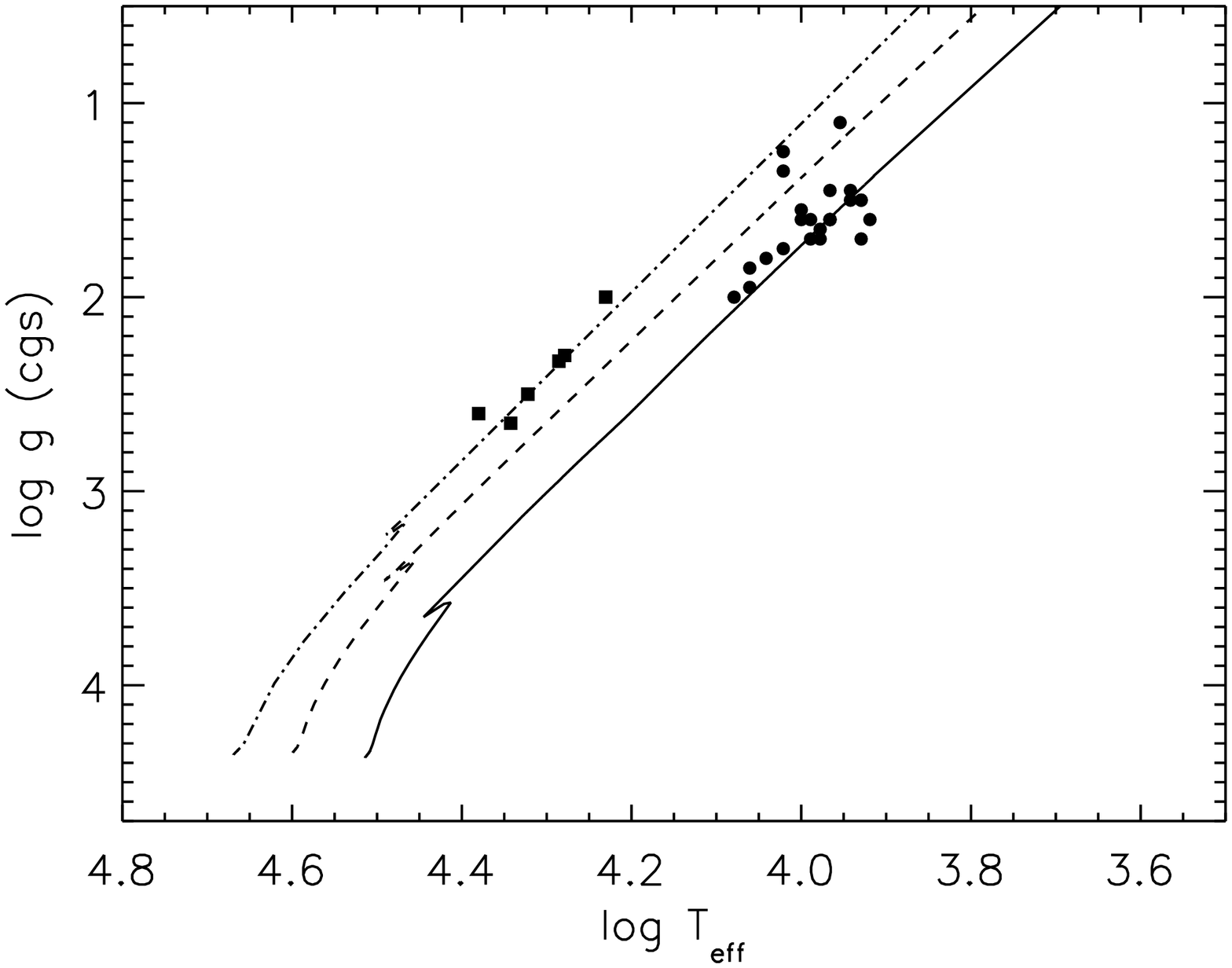}}
\end{minipage}
\hfill
\begin{minipage}{7cm}
   \resizebox{\hsize}{!}
   {\includegraphics[width=6cm,angle=90]{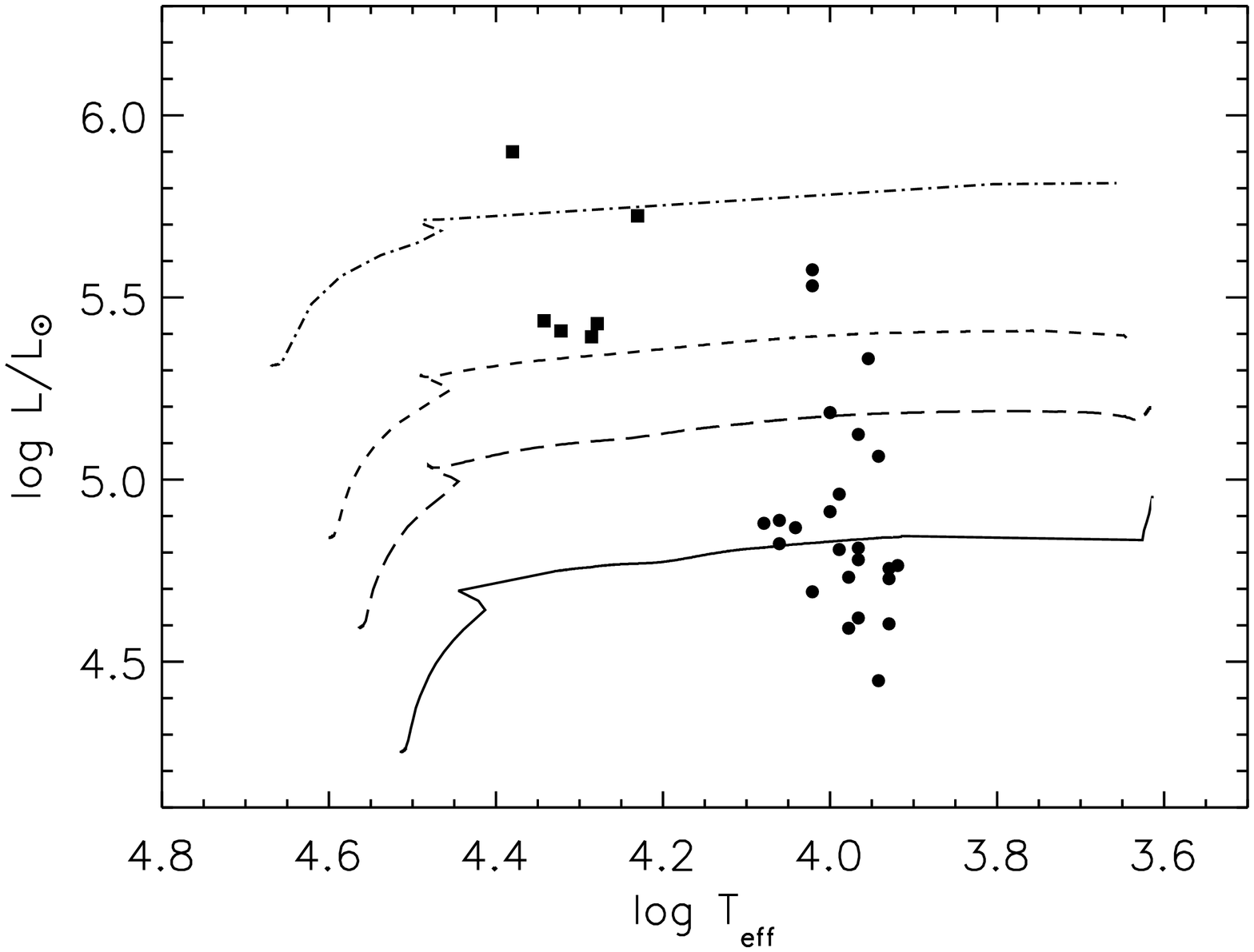}}
\end{minipage} 
\caption{{\bf Left:} NGC 300 A supergiants (filled circles) and early B supergiants (filled squares)
 in the $(log~g, log$~\teff) plane compared with evolutionary 
 tracks by \cite{meynet05} of stars with 15 M$_{\odot}$ (solid), 25 M$_{\odot}$ 
(dashed), and 40 M$_{\odot}$ (dashed-dotted), respectively.
   {\bf Right:} NGC 300 A and early B supergiants in the HRD compared with evolutionary 
 tracks for stars with 15 M$_{\odot}$ (solid), 
20 M$_{\odot}$ (long-dashed), 25 M$_{\odot}$ (short-dashed), 
and 40 M$_{\odot}$ (dashed-dotted), respectively. The tracks include the 
effects of rotation and are calculated for SMC metallicity (see 
\citealt{meynet05}).
}
   \label{fig5}
\end{figure}

\begin{figure}[b]
\begin{center}
 \includegraphics[width=8cm,angle=90]{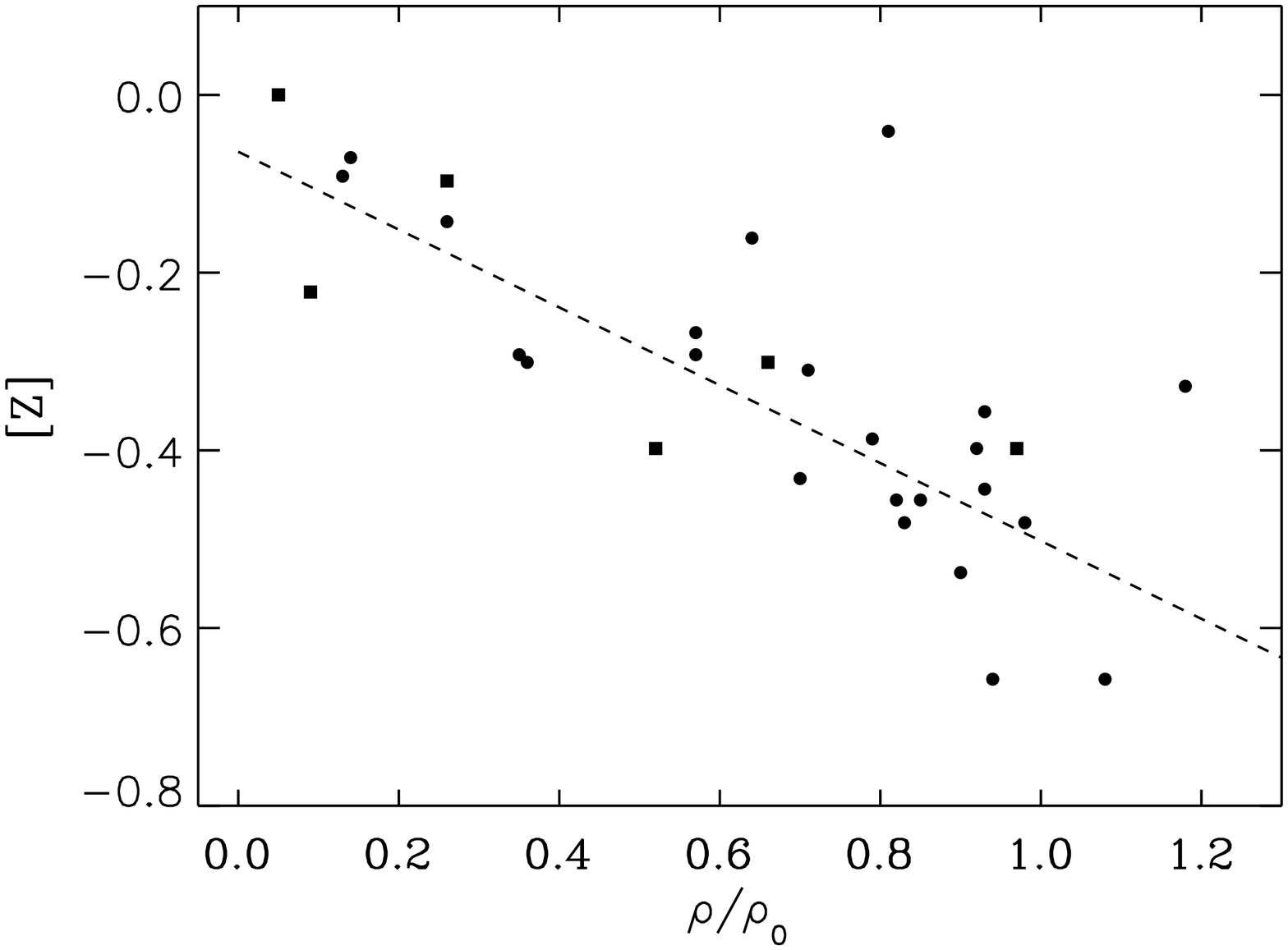} 
 \caption{Metallicity [Z] as a function of angular galacto-centric distance 
$\rho/\rho_{0}$ for the A supergiants (filled circles) and the 
early B-supergiants studied by \citet{urbaneja05} (filled squares). Note 
that for the latter metallicity refers to oxygen only. The dashed curve 
represents the regression discussed in the text.
}
   \label{fig6}
\end{center}
\end{figure}

\section{ A Pilot Study in NGC300 - Results}

As a first result, the quantitative spectroscopic method yields interstellar 
reddening and extinction as a by-product of the analysis process. For objects 
embedded in the dusty disk of a star forming spiral galaxy one expects a wide 
range of interstellar reddening E(B-V) and, indeed, a range from E(B-V) = 
0.07 mag up to 0.24 mag was found. The individual reddening values are 
significantly larger than the value of 0.03 mag 
adopted in the HST distance scale key project study of Cepheids by \cite{freedman01}
and demonstrate the need for a reliable reddening
determination for stellar distance indicators, at least as long the study is 
restricted to optical wavelengths. The average over the observed sample is 
$\langle E(B-V) \rangle $ = 0.12 mag in close agreement with the value of 0.1 mag 
found by \cite{gieren05} in their optical to near-IR study of Cepheids in NGC 300. 
While Cepheids have somewhat lower masses than the A supergiants of our study and 
are consequently somewhat older, they nonetheless belong to the same population and 
are found at similar sites. Thus, one expects them to be affected by interstellar 
reddening in the same way as A supergiants. 

Fig.\,\ref{fig5} shows the location of all the observed targets in the 
$(log~g, log$~\teff) plane and in the HRD compared to the early B-supergiants studied by
\cite{urbaneja05}. The comparison with evolutionary tracks gives a first 
indication of the stellar masses in a range from 10 M$_{\odot}$ to 40 M$_{\odot}$.
Three targets have obviously higher masses than the rest of the sample and seem to be
on a similar evolutionary track as the objects studied by \cite{urbaneja05}.
The evolutionary information obtained from the two diagrams appears to be 
consistent. The B-supergiants seem to be more massive than most of the A 
supergiants. The same three A supergiants apparently more massive than the rest 
because of their lower gravities are also the most luminous objects. 
This confirms that quantitative spectroscopy is -at least qualitatively - capable 
to retrieve the information about absolute luminosities. Note that the fact that 
all the B supergiants studied by \cite{urbaneja05} are more massive is simply 
a selection effect of the V magnitude limited spectroscopic survey by 
\cite{bresolin02}. At similar V magnitude as the A supergiants those objects 
have higher bolometric corrections because of their higher effective temperatures 
and are, therefore, more luminous and massive.  

Fig.\,\ref{fig6} shows the stellar metallicities and the metallicity gradient as
a function of angular galactocentric distance, expressed in terms of the isophotal radius, 
$\rho/\rho_{0}$ ($\rho_{0}$ corresponds to 5.33 kpc). Despite the scatter 
caused by the metallicity uncertainties of the 
individual stars the metallicity gradient of the young disk population in NGC 300 
is very clearly visible. A linear regression for the combined A- and B-supergiant 
sample yields (d in kpc)

\begin{equation}
        [Z] = -0.06\pm{0.09} - (0.083\pm{0.022})~d.
   \end{equation}

Note that the metallicities of the B supergiants 
refer to oxygen only with a value of log N(O)/N(H) = -3.31 adopted for the sun 
(\cite{allende01}). On the other hand, the A supergiant metallicities reflect 
the abundances of a variety of heavy elements such as Ti, Fe, Cr, Si, S, and Mg.

With these results KUBGP extended the discussion started by \cite{urbaneja05} 
to compare with oxygen abundances obtained from HII-region emission lines. 
\cite{urbaneja05} used line fluxes published by \cite{deharveng88} and applied 
various different published strong line method calibrations to determine nebular
oxygen abundances, which could then be used to obtain the similar regressions as 
above. 

The different strong line method calibrations lead to significant differences in 
the central metallicity as well as in the abundance gradient. The calibrations 
by \cite{dopita86} and \cite{zaritsky94} predict a metallicity significantly 
supersolar in the center of NGC 300 contrary to the other calibrations. On the other hand, the work by KUBGP 
yields a central metallicity slightly smaller than solar in good agreement with 
\cite{denicolo02} and marginally agreeing with \cite{kobulnicky99}, 
\cite{pilyugin01}, and \cite{pettini04}. At the isophotal radius, 5.3 kpc 
away from the center of NGC 300, they obtain an average metallicity significantly 
smaller than solar [Z] = -0.50, close to the average metallicity in the SMC. The 
calibrations by \cite{dopita86}, \cite{zaritsky94}, \cite{kobulnicky99} 
do not reach these small values for oxygen in the HII regions either because their 
central metallicity values are too high or the metallicity are gradients too shallow.

In light of the substantial range of metallicies obtained from HII region emission 
lines using different strong line method calibrations it seems to be extremely 
valuable to have an independent method using young stars. It will be very important 
to compare our results with advanced work on HII regions, which will avoid strong 
line methods, and will use direct infomation about nebular electron temperatures 
and densities (Bresolin et al., 2008, in preparation, see also this volume).

KUBGP discuss the few outliers in Fig.\,\ref{fig6} and claim that these metallicities 
seem to be real. Their argument is that the expectation of homogeneous azimuthal 
metallicity in patchy star forming galaxies seems to be naive. Future work on other 
galaxies will show whether cases like this are common or not.

\begin{figure}[b]
\begin{minipage}{7cm}
\resizebox{\hsize}{!}
   {\includegraphics[width=6cm,angle=90]{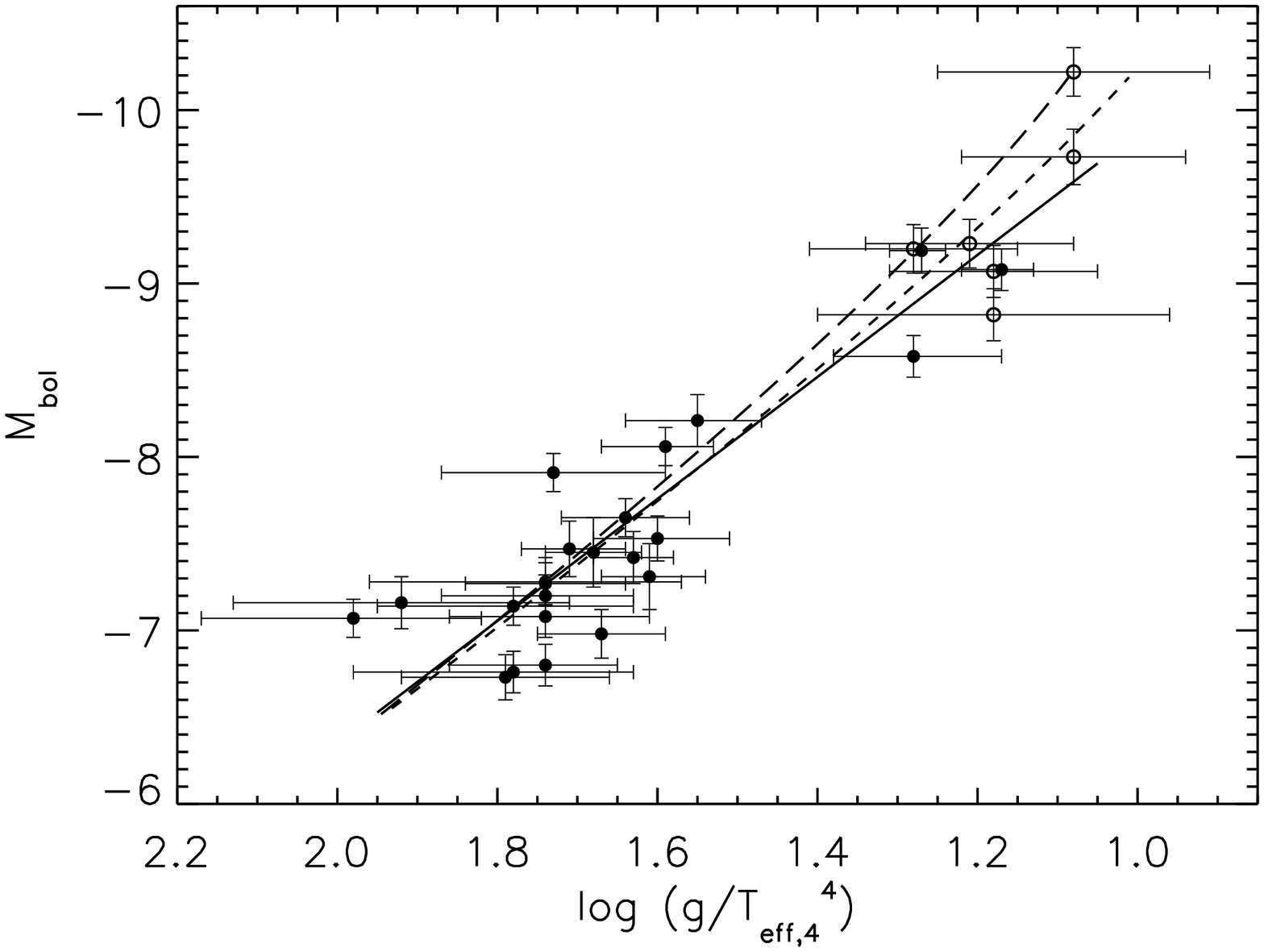}}
\end{minipage}
\hfill
\begin{minipage}{7cm}
   \resizebox{\hsize}{!}
   {\includegraphics[width=6cm,angle=90]{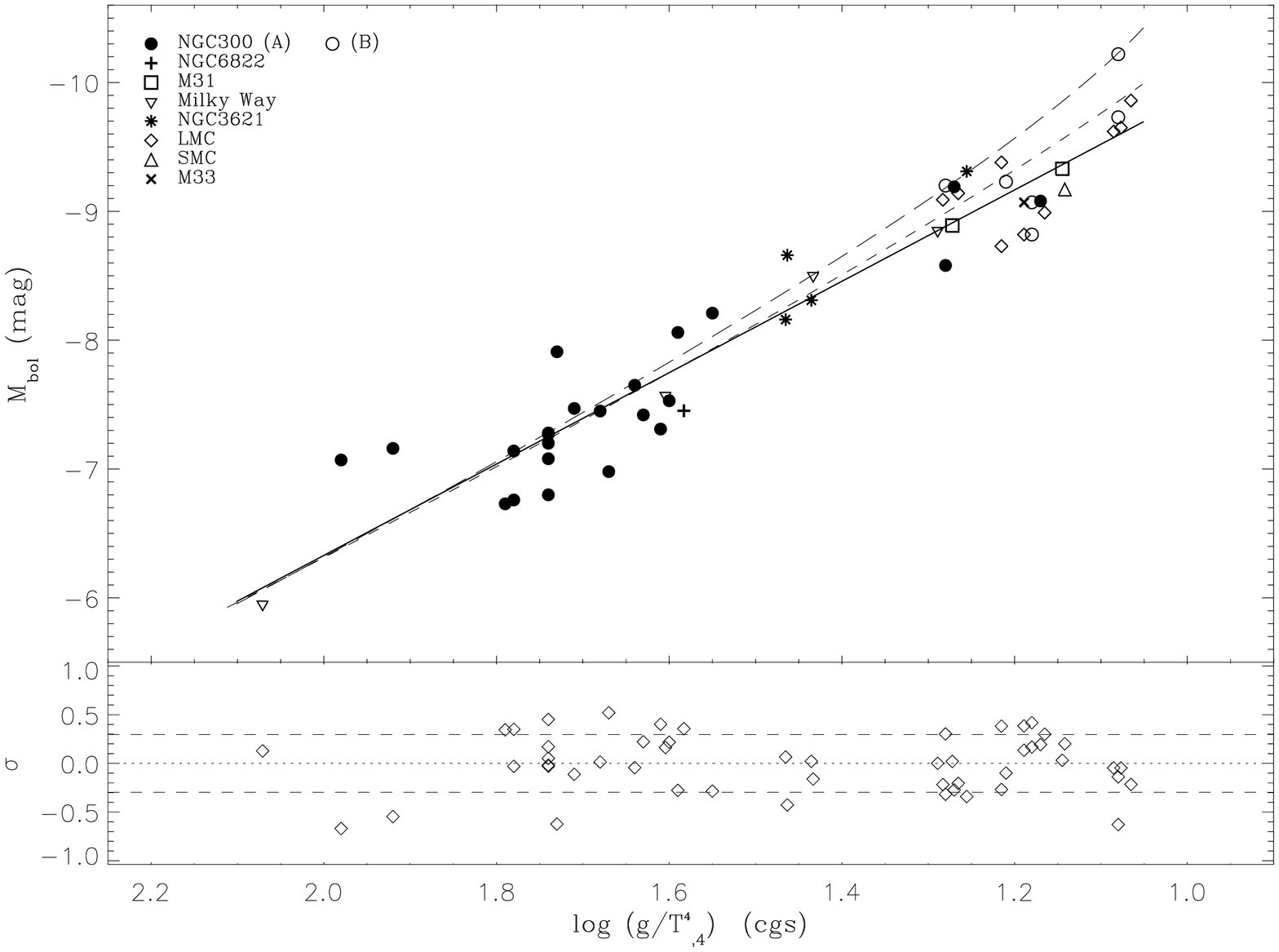}}
\end{minipage} 
\caption{{\bf Left:}
The FGLR of A (solid circles) and B (open circles) supergiants in 
NGC 300 and the linear regression (solid). The stellar evolution 
FGLRs for models with rotation are also overplotted 
(dashed: Milky Way metallicity, long-dashed: SMC metallicity).
   {\bf Right:} 
Same as left, but with additional objects from 7
other galaxies (see text).
}
   \label{fig7}
\end{figure}

\section{Flux Weighted Gravity - Luminosity Relationship (FGLR)}

Massive stars with masses in the range from 
12 M$_{\odot}$ to 40 M$_{\odot}$ evolve through the B and A 
supergiant stage at roughly constant luminosity (see Fig.\,\ref{fig5}). In addition, 
since the evolutionary timescale is very short when crossing through the B and A 
supergiant domain, the amount of mass lost in this stage is small. This means that 
the evolution proceeds at constant mass and constant luminosity. This has a very 
simple, but very important consequence for the relationship of gravity and 
effective temperature along each evolutionary track. From

\begin{equation}
L \propto R^{2}T^{4}_\mathrm{eff} = \mathrm{const.} ; M = \mathrm{const.}
\end{equation}
follows immediately that
\begin{equation}
M \propto g\;R^{2} \propto L\;(g/T^{4}_\mathrm{eff}) = L~g_{F} = \mathrm{const.}
\end{equation}

Thus, along the evolution through the B and A supergiant domain the 
\emph{``flux-weighted gravity''} $g_{F} = g$/\teffq\/ should remain constant. 
This means each evolutionary track of different luminosity in this domain is characterized 
by a specific value of $g_{F}$. This value is determined by the relationship 
between stellar mass and luminosity, which in a first approximation is a power law
\begin{equation}
L \propto M^{x}\;
\end{equation}

and leads to a realtionship between luminosity and flux-weighted gravity

\begin{equation}
L^{1-x} \propto (g/T^{4}_\mathrm{eff})^{x}\;.
\end{equation}

With the definition of bolometric magnitude 
$M_\mathrm{bol}$\,$\propto$\,$-2.5\log L$ one then derives

\begin{equation}
-M_\mathrm{bol} = a_{FGLR}(\log~g_{F} - 1.5)) + b_{FGLR}\;.
\end{equation}

This is the  \emph{``flux-weighted gravity -- luminosity relationship''} 
(FGLR) of blue supergiants. Note that the proportionality constant
$a_{FGLR}$ is given by the exponent of the mass -- luminosity power law through

\begin{equation}
a_{FGLR} = 2.5 x/(1-x)\;,
\end{equation}

for instance, for $x=3$, one obtains $a_{FGLR}=-3.75$. Note that the zero 
point of the relationship is chosen at a flux weighted gravity of 1.5, which is in the 
middle of the range encountered for blue supergiant stars.

KUBGP use the mass-luminosity relationships of different evolutionary tracks
(with and without rotation, for Milky Way and SMC metallicity) to calculate the 
different FGLRs predicted by stellar evolution. Very interestingly, 
while different evolutionary model types yield somewhat different FGLRs, 
the differences are rather small. 

\cite{kud03} were the first to realize that the FGLR has a very interesting 
potential as a purely spectroscopic distance indicator, as it relates two 
spectroscopically well defined quantitities, effective temperature and gravity, 
to the absolute magnitude. Compiling a large data set of spectroscopic high 
resolution studies of A supergiants in the Local Group and with an approximate 
analysis of low resolution data of a few targets in galaxies beyond the Local 
Group (see discussion in previous chapters) they were able to prove the 
existence of an observational FGLR rather similar to the theoretically 
predicted one.

With the improved analysis technique of low resolution spectra of A supergiants 
and with the much larger sample studied for NGC 300 KUBGP resumed the 
investigation of the FGLR.

 The result is shown in Fig.\,\ref{fig7}, which for NGC 300 (left diagram) reveals 
a clear and rather 
tight relationship of flux weighted gravity $log~g_{F}$ with bolometric magnitude 
$M_{bol}$. A simple linear regression yields $b_{FGLR}$ = 8.11 for the zero point 
and 
$a_{FGLR}$ = -3.52 for the slope. The standard deviation from this relationship is 
$\sigma$ = 0.34 mag. Within the uncertainties the observed FGLR appears to be in 
agreement with the theory.

In their first investigation of the empirical FGLR \cite{kud03} have added A supergiants 
from six Local Group galaxies with stellar parameters obtained from quantitative studies 
of high resolution spectra (Milky Way, LMC, SMC, M31, M33, NGC 6822) to their results 
for NGC 300 to 
obtain a larger sample. They also added 4 objects from the spiral galaxy NGC 3621 
(at 6.7 Mpc) which were studied at low resolution. KUBGP added exactly the same 
data set to their new enlarged NGC 300 sample, however, with a few minor modifications. 
For the Milky Way they included the latest results from \cite{przybilla06} and 
\cite{schiller08} and 
for the two objects in M31 we use the new stellar parameters obtained by 
\cite{przybilla06b}. For the objects in NGC 3621 they applied new HST photometry. 
They also re-analyzed the LMC objects using ionization equilibria for the 
temperature determination.

Fig.\,\ref{fig7} (right diagram) shows bolometric magnitudes and flux-weighted 
gravities for this 
full sample of eight galaxies revealing a tight relationship over one order of 
magnitude in flux-weighted gravity. The linear regression coefficients are 
$a_{FGLR} = -3.41\pm{0.16}$ and $b_{FGLR} = 8.02\pm{0.04}$, very simimilar to 
the NGC 300 sample 
alone. The standard deviation is $\sigma$ = 0.32 mag. The stellar evolution FGLR 
for Milky Way metallicity 
provides a fit of almost similar quality with a standard deviation of 
$\sigma$ = 0.31 mag.

\section{Conclusions and Future Work}

The astrophysical potential of low resolution spectroscopy of A supergiant stars 
for studying galaxies beyond the Local Group is quite remarkable.
By introducing a novel method for the quantitative spectral analysis one is 
able to determine accurate stellar parameters, which allow for a detailed test of 
stellar evolution models. Through the spectroscopic determination of stellar 
parameters one can also 
constrain interstellar reddening and extinction by comparing the calculated SED 
with broad band photometry. The study of NGC 300 finds a very patchy extintion pattern as to be 
expected for a star forming spiral galaxy. The average extinction is in agreement 
with multi-wavelength studies of Cepheids including K-band photometry.

The method also allows to determine stellar metallicities and to study stellar 
metallicity gradients. Solar metallicity is found in the center of 
NGC 300 and a gradient of $-$0.08 dex/kpc. To our knowledge this is the first 
systematic stellar metallicity study in galaxies beyond the Local Group 
focussing on iron group elements. In the future the method can be extended to 
not only determine metallicity but also the ratio of $\alpha$- to iron group 
elements as a function of galactocentric distance. The stellar metallicities 
obtained can be compared with oxygen abundance studies of HII regions using 
the strong line method. This allows to discuss the various calibrations 
of the strong line method, which usually yield very different results.

The improved spectral diagnostic method makes it possible to very 
accurately determine stellar flux weighted gravities $log~g_{F}$ = $log~g$/\teffq\/ 
and bolometric magnitudes. Above a 
certain threshold in effective temperature a simple measurement of the strengths 
of the Balmer lines can be used to determine accurate values of $log~g_{F}$.

Absolute bolometric magnitudes $M_{bol}$ and flux-weighted gravities $log~g_{F}$ 
are tightly correlated. It is shown that such a correlation is expected for 
stars, which evolve at constant luminosity and mass. This 
``flux-weighted gravity - luminosity relationship (FGLR)'' agreement with 
stellar evolution theory.

With a relatively small residual scatter of $\sigma$ = 0.3 mag the observed 
FGLR is an excellent tool to determine accurate spectroscopic distance to galaxies.
It requires multicolor photometry and low resolution ($5\AA$) spectroscopy to
determine effective temperature and gravity and, thus, flux-weighed gravity 
directly from the spectrum.
With effective temperature, gravity and metallicity determined one also knows the 
bolometric correction, which is small for A supergiants, which means that errors 
in the stellar parameters do not largely affect the determination of bolometric 
magnitudes. Moreover, one knows the intrinsic stellar SED and, therefore, can 
determine interstellar reddening and extinction from the multicolor photometry, 
which then allows for the accurate determination of the reddening-free apparent 
bolometric magnitude. The application of the FGLR then yields absolute 
magnitudes and, thus, the distance modulus. With the intrinsic scatter of 
$\sigma$ = 0.3 mag and 30 targets per galaxy one can estimate an accuracy of 
0.05 mag in distance modulus (0.1 mag for 10 target stars). The results for WLM 
by Miguel Urbaneja et al. and for M33 by Vivian U et al. presented in these 
proceedings are a first demonstration of the power of the method.

The advantage of the FGLR method for distance determinations is its 
spectroscopic nature, which provides significantly more information about 
the physical status of the objects used for the distance determination than 
simple photometry methods. Most importantly, metallicity and interstellar 
extinction can be determined directly. The latter is crucial for spiral and 
irregular galaxies because of the intrinsic patchiness of reddening and 
extinction.

Since supergiant stars are known to show intrinsic photometric variability, 
the question arises whether the FGLR method is affected by such variability. 
For the targets in NGC 300 this issue has been carefully investigated 
by \citet{bresolin04}, who studied CCD photometry lightcurves 
obtained over many epochs in the parallel search for Cepheids in NGC 300. They 
concluded that amplitudes of photometric variability are very small and do 
not affect distance determinations using the FGLR method.

The effects of crowding and stellar multiplicity are also important. However,
in this regard A supergiants offer tremendous advantages relative to other 
stellar distance indicators. First of all, they are significantly brighter.
\citet{bresolin05} using HST ACS photometry compared to ground-based 
photometry have studied the effects of crowding on the Cepheid distance 
to NGC 300 and concluded that they are negligible. With A supergiants being 
3 to 6 magnitudes brighter than Cepheids it is clear that even with 
ground-based photometry only crowding is generally not an issue for these 
objects at the distance of NGC 300 and, of course, with HST photometry 
(and in the future JWST) one can reach much larger distances before crowding 
becomes important. In addition, any significant contribution by additional 
objects to the light of an A supergiant will become apparent in the spectrum, 
if the contaminators are not of a very similar spectral type, which is very 
unlikely because of the short evolutionary lifetime in the A supergiant stage. 
It is also important to note that A supergiants have evolutionary ages larger 
than 10 million years, which means that they have time to migrate into the 
field or that they are found in older clusters, which are usually less concentrated
than the very young OB associations.

It is evident that the type of work described in this paper can be 
in a straightforward way extended to the many spiral galaxies in the local volume 
at distances in the 4 to 7 Mpc range. Pushing the method we estimate 
that with present day 8m to 10m class telescopes and the existing very efficient 
multi-object spectrographs one can reach down with sufficient S/N to V = 22.5 mag in two nights of 
observing time under very good conditions. For objects brighter than 
$M_{V}$ = -8 mag this means metallicities and distances can be determined out 
to distances of 12 Mpc (m-M = 30.5 mag). This opens up a substantial volume 
of the local universe for metallicity and galactic evolution studies and 
independent distance determinations complementary to the existing methods. With 
the next generation of extremely large telescopes such as the TMT, GMT or the 
E-ELT the limiting magnitude can be pushed to V = 24.5 equivalent to 
distances of 30 Mpc (m-M = 32.5 mag).

\begin{discussion}

\discuss{Massey}{This is absolutely beautiful work! Let me ask one quesition, 
though! You have described the precision with which you can measure effective 
temperatures, surface gravities, abundances, etc. Do you have a sense of their 
systematic accuracy? Your models must be the best possible now, but what might 
affect the results of the model that you are unsure of?}

\discuss{Kudritzki}{There is always the possibility of systematic effects 
caused by the imperfectness of the models and the NLTE radiative transfer. 
There are several ways how one can try to address those. For the effective 
temperatures one can use different methods, ionization equilibria and Balmer 
jumps, for instance. When the results agree, this is a good indication. With 
our IfA student Ben Granett we have also tried the alternative approach to 
determine \teff from the full integral over the complete UV, optical, IR energy
distribution and we found good agreement again. You can test gravities with 
detached eclipsing binaries. This has been done in the most recent paper by 
Bonanos et al., 2006. For abundances, the rule is to use as many lines as 
possible, also in different spectral windows, and to look at the scatter. 
This work has been done very carefully by Przybilla and collaborators 
(see his poster papers at this conference).}

\discuss{Burbidge}{I wish to make three points. (1) This is marvellous work.
The spectroscopic method of distance determination is very powerful. (2) Don't 
be convinced by those who claim that the Hubble Constant is well determined. 
The value obtained by Sandage and Tammann of about 55 km/sec/Mpc is much more 
likely correct than the value of 71 km/sec/Mpc being claimed. (3) How long will 
it take to extend the method to the Virgo Cluster?}

\discuss{Kudritzki}{Many thanks. I do not have a bias towards any value of the 
Hubble constant. I think that it is important to have as many independent 
reliable methods as possible. That is the reason why I am convinced that the 
FGLR-method is really important. It is an independent way to check existing 
results. This is also the philosphy of the Araucaria-Project, where we compare 
Cepheid-distances with TRGB-distances and with the FGLR-method. As for the Virgo 
cluster, this is a real challenge with Keck, but it will be easy with the TMT.}

\discuss{D'Odorico}{With ELTs like the TMT you expect to achieve the maximum 
angular resolution at IR wavelengths, where you can use AO at best. Do you think 
it would be possible to do this type of studies of the most luminous supergiant 
stars based on IR spectra?}

\discuss{Kudritzki}{A lot of work has been done already on the IR quantitative 
spectroscopy of supergiants, as for instance presentated by the papers by 
Paco Najarro and Fabrice Martins at this meeting. Norbert Przybilla has done 
a lot already on A supergiants. It will certainly be worthwhile to extent this 
work to the IR, although the B and A supergiants emit less flux in this spectral 
window. At the same time, I think, it will also be important to push the AO limit 
as much towards the visual as by any means possible. While this will certainly 
not be possible for the ELT first light instruments, in is not excluded in the 
long term.}

\end{discussion}

\end{document}